# A Detailed Analysis and Monte Carlo Simulation of the Neutron Lifetime Experiment S. Arzumanov et al., Phys. Lett. B 483 (2000) 15

A.K. Fomin, A.P. Serebrov[*]

*Petersburg Nuclear Physics Institute, Russian Academy of Sciences, Gatchina, Leningrad District, 188300, Russia*

[*] *Corresponding author:* A.P. Serebrov

A.P. Serebrov

Petersburg Nuclear Physics Institute

Gatchina, Leningrad district

188300 Russia

Telephone: +7 81371 46001

Fax: +7 81371 30072

E-mail: serebrov@pnpi.spb.ru

**Abstract**

We performed a detailed analysis and the Monte Carlo simulation of the neutron lifetime experiment [1] because of the strong disagreement by 5.6 standard deviations between the results of this experiment and our experiment [2]. We found a few effects which were not taken into account in the experiment [1]. The possible correction is −5.5 s with uncertainty of 2.4 s which comes from initial data knowledge. We assume that after taking into account this correction the result of work [1] for neutron lifetime 885.4 ± 0.9$_{stat}$ ± 0.4$_{syst}$ s could be corrected to 879.9 ± 0.9$_{stat}$ ± 2.4$_{syst}$ s.

**Introduction**

The recent neutron lifetime experiment [2] has provided the value 878.5 ± 0.8 s. It differs by 6.5 standard deviations from the world average value 885.7 ± 0.8 s quoted by the particle data group (PDG) in 2006 [3] and by 5.6 standard deviations from the previous most precise result 885.4 ± 0.9$_{stat}$ ± 0.4$_{syst}$ s [1]. Our experiment employed a gravitational trap with low-temperature fluorinated oil (fomblin) coating, which provides several advantages with respect to previous experiments. First of all, a small loss factor of only 2·10$^{-6}$ per collision of UCN with trap walls results in a low loss probability of only 1% of the probability of neutron β-decay. Therefore the measurement of neutron lifetime was almost direct; the extrapolation from the best storage time to the neutron lifetime was only 5 s. In these conditions it is practically impossible to obtain a systematical error of about 7 s. The quoted systematical error of the experimental result [2] was 0.3 s.

In determination of the world average value of the neutron lifetime there is rather dramatic situation. On the one hand a new value of the neutron lifetime from work [2] cannot be included in the world average value because of a big difference of results. On the other hand until this major disagreement is understood the present world average value for the neutron lifetime must be suspect. The situation on PDG page devoted to the neutron lifetime is formulated in view of this particular controversy.

The only way out of the present situation is to carry out new more precise experiments. A more detailed analysis of previous experiments and search of possible systematic errors are also reasonable. We cannot find by any means an error in 7 s in our measurements [2] where extrapolation of UCN storage time to the neutron lifetime is only 5 s. Therefore we have made the analysis of experiment [1] where extrapolation is 100-120 s and at the same time it is affirmed that it is done with systematic error 0.4 s. It is this state of things that causes obvious doubts. The detailed analysis of experiment [1] carried on by Monte Carlo simulation is made below.

**Scheme and method of the experiment [1]**

Below we reproduce a short description of the experiment [1] using mainly the text of the article. The setup is shown in Fig. 1. The storage vessel (7), (8) is composed of two coaxial horizontal cylinders made of aluminium of 2 mm thickness. The cylinder walls were coated with a thin layer of Fomblin oil which has very low UCN losses. In order to maintain this oil layer on the surface, the cylinder walls were first coated by a layer of Fomblin grease of about 0.2 mm thickness.

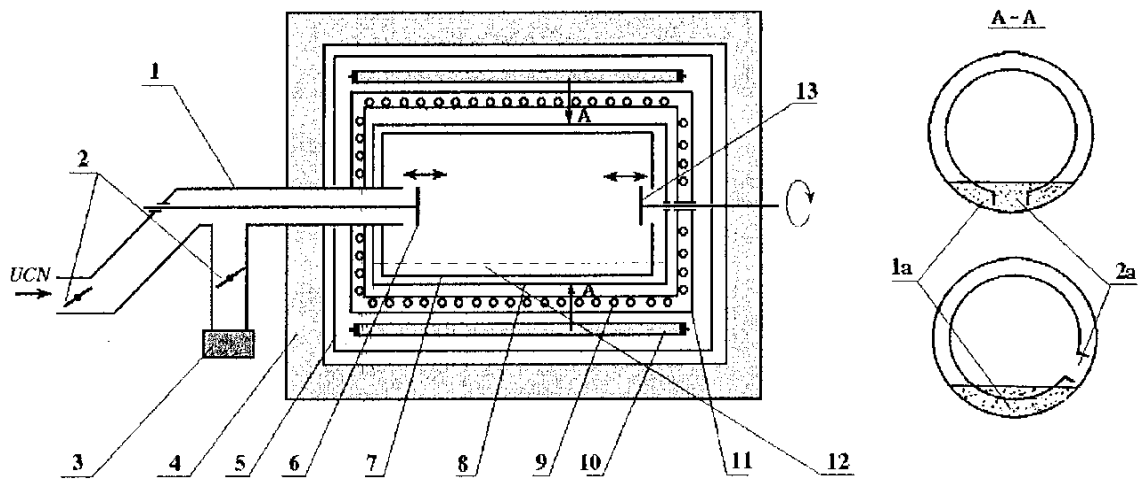

Fig. 1. The scheme of the experimental setup. 1 - UCN guide, 2 - shutters, 3 - UCN detector, 4 - polyethylene shielding, 5 - cadmium housing, 6 - entrance shutter of the inner vessel, 7 - inner storage vessel, 8 - outer storage vessel, 9 - cooling coil, 10 - thermal neutron detector, 11- vacuum housing, 12 - oil puddle, 13 - entrance shutter of the gap vessel, 1a - oil puddle, 2a - slit.

The inner cylinder (7) was 33 cm in diameter and 90 cm long, while the dimensions of the outer one (8) were larger by a gap of 2.5 cm. The shutter (6) connects the inner cylinder to the intermediate chamber which has connections (i) to the neutron guide (1) of the TGV UCN source by the entrance shutter and (ii) to the UCN detector (3) by shutter (2). The shutter (13) connects the inner cylinder to the volume of the annular gap between both cylinders.

The inner cylinder had a long slit (2a) of a special form (see Fig. 1) along a cylinder surface. The edges of the slit were dipped into a Fomblin oil puddle (1a) with level (12) when the slit was situated at the bottom position during storage. The construction allowed to rotate the cylinders in common about its horizontal axis without a vacuum break to refresh the oil layers on the cylinder walls.

The storage vessel was placed inside the vacuum housing (11). The vessel volume was hermetically sealed from the housing. The housing was formed by two coaxial cylinders of stainless steel. The outer surface of the inner cylinder had a serpent tube (9) to cool the bottles. The cooling system stabilized the bottle temperature which could be set in the range +20°C ÷ − 26°C.

The set-up was surrounded by the thermal neutron detectors comprising a set of 24 counters of the SNM-57 type (10), each counter being a $^3$He filled tube of 3 cm diameter and 100 cm long. The UCN detector was a $^3$He loaded proportional counter (3) with an Al entrance window of 100 μm thickness.

The whole installation was placed inside the shielding (5) of 1 mm thick Cd and the shielding (4) of 16 cm thick boron polyethylene.

The construction permitted to store UCN either in the inner cylinder or in the annular space between the inner and outer cylinder, thereby changing the UCN loss rate by a factor of about 5 without breaking the vacuum.

The experiment was carried out using the following sequence of procedures.

1. **Filling.** The chosen vessel, annular or central, was filled for 200 s. For filling only the central vessel the shutter 13 was closed. For the annular vessel shutter 13 was open and the UCN removed from the central vessel in the following step.

2. **Cleaning.** The trapped neutron spectrum in the storage vessel was given time to clean during $t_{cl}$ (200 s to 1000 s). This procedure was necessary as the UCN source provided a rather broad neutron spectrum. During the cleaning time $t_{cl}$ UCN with velocity exceeding the limiting velocity of Fomblin escaped from the vessel. When the annular vessel was chosen the shutter 6 and the shutter to the UCN detector were opened during $t_{cl}$ to empty the central vessel.

3. **Emptying.** The UCN were emptied to the detector from the chosen vessel and counted for 200 s yielding the initial quantities $N_i$ and $n_i(t)$, where $n_i(t)$ denotes the counting rate in the UCN detector during the emptying time $t$ and $N_i$ the integral over $n_i(t)$. On emptying the inner vessel both its shutters were opened to make the emptying conditions more equal for the two vessels.

4. Steps 1. and 2. were repeated to fill the chosen vessel and to clean the UCN spectrum before the storage period. Due to the stable intensity of the UCN source the initial conditions were essentially identical.

5. **Storing.** After the cleaning time the UCN were further stored in the chosen vessel for the time T and the inelastically scattered and leaked neutrons were counted during that interval in the thermal neutron and UCN detector, respectively.

6. Recording of the final UCN quantity $N_f$ and $n_f(t)$ by counting for 200 s (same procedure as step 3).

7. The background of the detectors was measured during 150 s after all UCN have left the vessel.

All abovementioned procedures of the experiment are shown in Fig. 2.

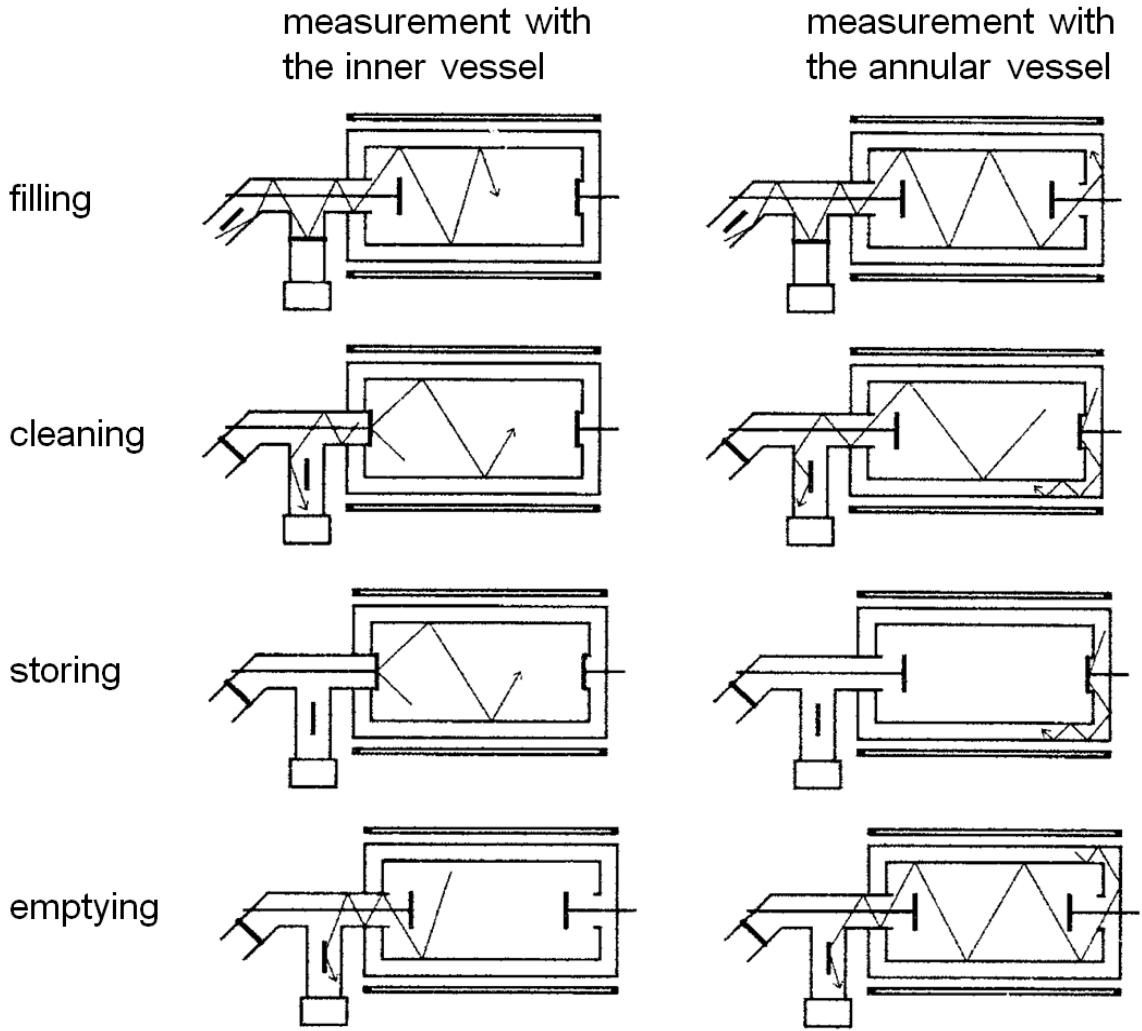

Fig. 2. The procedures of the experiment.

Basic idea of the experimental method for a monoenergetic UCN spectrum is the following. The number of neutrons $N(t)$ in the trap changes exponentially during the storage time, i.e. $N(t) = N_0 e^{-\lambda t}$. The value $\lambda$ is the total probability per unit time for the disappearance of UCN due to both the beta-decay and losses during UCN-wall collisions. In turn, losses are equal to the sum of the inelastic scattering rate constant $\lambda_{ie}$, and that for the neutron capture at the wall, $\lambda_{cap}$:

$$\lambda = \lambda_n + \lambda_{loss} = \lambda_n + \lambda_{ie} + \lambda_{cap} \qquad (1)$$

The ratio $\lambda_{cap}/\lambda_{ie}$ is to a good approximation equal to the ratio of the UCN capture and inelastic scattering cross sections for the material of the wall surface since both values are proportional to the wall reflection rate of UCN in the trap. Hence $\sigma_{cap}/\sigma_{ie}$ and the value

$$a = \lambda_{loss}/\lambda_{ie} = 1 + \lambda_{cap}/\lambda_{ie} = 1 + \sigma_{cap}/\sigma_{ie} \qquad (2)$$

is constant for the given conditions, i.e. same wall material and temperature. During storage the upscattered neutrons are recorded with an efficiency $\varepsilon_{th}$ in the thermal neutron detector surrounding the storage trap. The corresponding counting rate is given by

$$j = \varepsilon_{th}\lambda_{ie}N(t) \qquad (3)$$

Hence the total counts in the time interval $T$ are equal to

$$J = \varepsilon_{th}\lambda_{ie}(N_0 - N_T)/\lambda \qquad (4)$$

Here $N_0$ and $N_T$ are the UCN populations in the trap at the beginning and the end of the storage time $T$, respectively. The UCN themselves are measured with an efficiency $\varepsilon$ such that the detected UCN at the beginning (normalisation measurement) and the end of the storage time are equal to $N_i = \varepsilon N_0$ and $N_f = \varepsilon N_T$ respectively. We have then

$$\lambda_{ie} = \frac{J\lambda}{N_i - N_f}\frac{\varepsilon}{\varepsilon_{th}} \qquad (5)$$

$$\lambda = \frac{1}{T}\ln(N_i/N_f) \qquad (6)$$

The experiment is repeated with a different value for the wall loss rates. The ratio of the two corresponding $\lambda$ values are built following Eq. (1) and including Eq. (2) with constant value $a$. Thus $\lambda_n$ is given by

$$\lambda_n = \frac{\xi\lambda^{(1)} - \lambda^{(2)}}{\xi - 1} \qquad (7)$$

where

$$\xi = \lambda_{ie}^{(2)}/\lambda_{ie}^{(1)} \qquad (8)$$

The indices refer to the two measurements with different $\lambda_{loss}$. The expression Eq. (7), (8) contains then only the directly measured quantities $J$, $N_i$, $N_f$ following Eqs.

(5), (6) since the efficiencies of the neutron detection cancel. The value for $\lambda_{loss}$ can be varied by changing the ratio of the surface to the volume of the bottle and hence the reflection rate with the walls. In order to keep the value $a$ constant the (monoenergetic) energy of the UCN and the specification of the wall (temperature, type of wall, etc.) must be the same.

Description of this method for a broad UCN spectrum and more experimental details can be found in [1].

**The Analysis and Monte Carlo Simulation of the experiment [1]**

Processing of results of a method of work [1] for extrapolation to the neutron lifetime is presented by Eqs. (7), (8). For descriptive reasons (Fig. 3a) it is possible to suggest the graphic solution, using Eqs. (1), (2).

From Eqs. (1), (2) we can write:

$$\lambda = \lambda_n + a\lambda_{ie}. \tag{9}$$

Accordingly for two measurements in different geometry:

$$\lambda^{(1)} = \lambda_n + a\lambda_{ie}^{(1)}, \tag{10}$$

$$\lambda^{(2)} = \lambda_n + a\lambda_{ie}^{(2)}. \tag{11}$$

Excluding $a$ from the system of equations:

$$\lambda_n = \frac{\lambda^{(1)}\lambda_{ie}^{(2)} - \lambda^{(2)}\lambda_{ie}^{(1)}}{\lambda_{ie}^{(2)} - \lambda_{ie}^{(1)}} = \frac{\xi\lambda^{(1)} - \lambda^{(2)}}{\xi - 1}, \tag{12}$$

where $\xi = \lambda_{ie}^{(2)} / \lambda_{ie}^{(1)}$, i.e. we derive Eq. (7) of work [1].

It is quite obvious that for absence of systematic in a method of work [1] it is necessary to have full equivalence of parameters $\lambda$ and $\lambda_{ie}$ for two different vessels. We will consider possible distinctions for $\lambda$ and $\lambda_{ie}$ which arise at change of geometry of experiment.

MC simulation of the experiment [1] was performed using a code capable of taking into account gravity. The code was written by A.K. Fomin especially for simulations with UCN. This code starts with an initial distribution of neutrons and calculates the track of each particle analytically until it reaches a material boundary. At each wall collision the loss and reflection probability is calculated, resulting in a new direction to calculate the trajectory until the next boundary is reached. The code uses specula and diffusion reflections with walls.

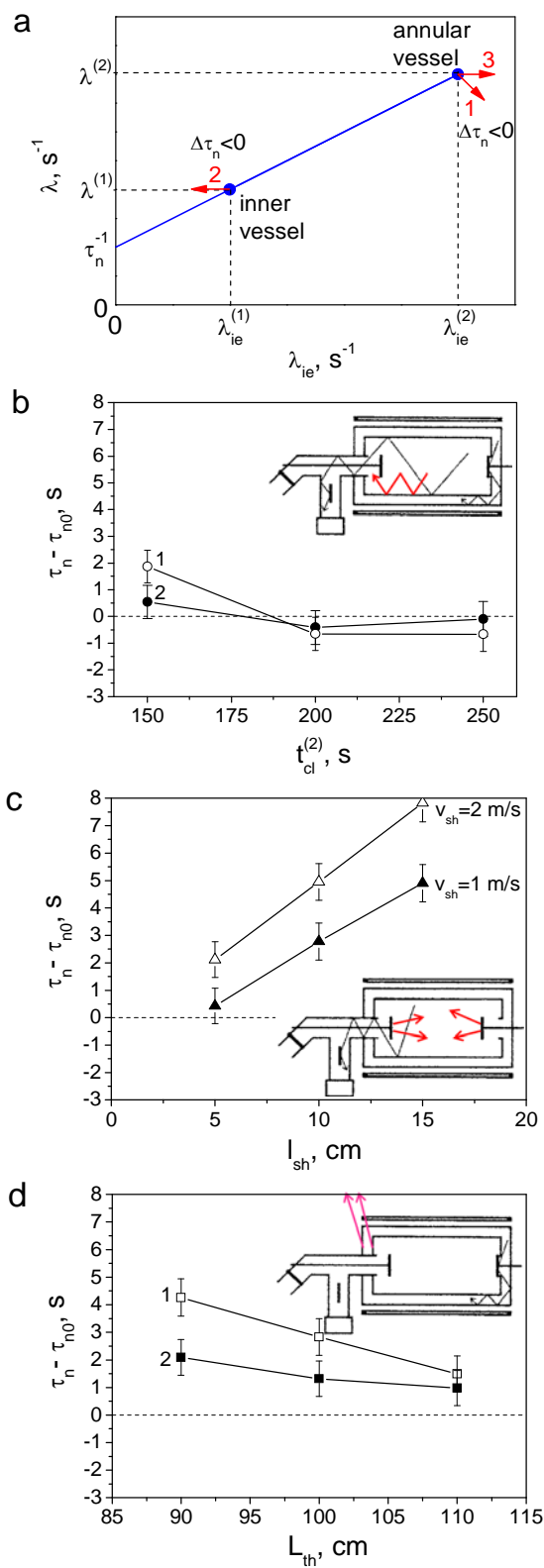

Fig. 3. (a) diagram showing influence of various effects for measured value of neutron lifetime; (b) correction of neutron lifetime due to effect of not full emptying of the inner vessel during cleaning while working with the annular vessel: simulations for neutron guide length in front of the detector of 0.8 m (curve 1) and 1 m (curve 2); (c) correction of neutron lifetime due to effect of heating of neutrons by the shutters; (d) correction of neutron lifetime due to effect of not equal thermal neutron detection efficiencies for different vessels: simulations without capture and scattering in materials (curve 1) and with capture and scattering in materials (curve 2).

The geometry of setup and time intervals were chosen the same as in the experiment. After each simulation we have values of $N_i$, $N_f$, $J$, $j(t)$ and $n(t)$. We evaluate the obtained data in the same way as in the experiment. In all our simulations neutron lifetime was fixed to a definite value. Repeating experimental procedure we obtain the extrapolated neutron lifetime values and comparing it with initial one we get correction to the experimental result.

The percentage values of diffusion reflections by walls were set to reproduce an experimental emptying process, i.e. time dependence of UCN detector count in the course of registration. It is the most detailed information which can be found in work [1]. Neutron reflection by walls was approximated by 50% specular and 50% diffusion reflections for the inner and annular vessels. Such a factor seems to be reasonable since the surface of vessels has been covered by a layer of Fomblin grease before being covered with Fomblin oil. Neutron reflection by walls was approximated by 90% specular and 10% diffusion reflections for the neutron guides. That corresponds to quality of electropolished neutron guides.

MC simulation was done for the temperature –26°C because most of the experimental data was obtained at this temperature.

We studied three effects in MC simulations: (1) not full emptying of the inner vessel during cleaning while working with the annular vessel; (2) heating of neutrons by shutters; (3) not equal thermal neutron detection efficiencies for different vessels.

**1. Effect of not full emptying of the inner vessel during cleaning while working with the annular vessel.** One can see from Fig. 2 that process of UCN emptying to the detector after holding in the inner and the annular vessels is different. Emptying after holding in the inner vessel occurs directly to the detector through neutron guide system. However, after holding in the annular vessel neutrons at first pass through the inner vessel. The authors of work [1] try to make conditions of emptying more identical and on emptying the inner vessel both its shutters were opened to make the emptying conditions more equal for the two vessels. The question arises how perfect emptying the inner vessel will be released before opening of the shutter 13 for emptying the annular vessel after cleaning. For an estimation of a possible systematic error in this process we have done MC simulation of the process taking into account geometry of experiment [1].

The shutter 6 and the shutter of UCN detector are opened during $t_{cl}$ when we work with the annular vessel. It is necessary to empty the inner vessel from UCN during holding in the annular vessel. If this time is not enough for the inner vessel there are still neutrons which are added to neutrons from the annular vessel during its emptying. It gives higher value of $N_i$ and correspondingly higher value of $\lambda$ and lower value of $\lambda_{ie}$ for the annular vessel:

$$\lambda_{ie} = \frac{J\lambda}{(N_i + \Delta N_i) - N_f} \frac{\varepsilon}{\varepsilon_{th}}, \qquad (13)$$

$$\lambda = \frac{1}{T} \ln\left((N_i + \Delta N_i)/N_f\right), \qquad (14)$$

where $\Delta N_i$ is number of UCN in the inner vessel after cleaning in the annular vessel. The arrow (1) in Fig. 3a shows the direction of changed position of point for the annular vessel after correction. It gives negative correction for measured value of neutron lifetime. The values of $t_{cl}$ for MC simulation are taken from Table 1 [1]. The results of extrapolations to neutron lifetime are shown in Fig. 3b for different $t_{cl}$ and different neutron guide length in front of the detector which has not been strictly defined in the data of geometry of experiment. By results of simulation it is possible to draw a conclusion that the effect of an incomplete emptying has not been found out, though uncertainty of an estimation of this process is at level of 1 s.

**2. Effect of heating of neutrons by shutters.** The following non-equivalence of measurements for different vessels is observed at emptying. Before release of neutrons to the detector the shutters 6 and 13 are open. At shutter movement in volume with UCN there is either heating or cooling of UCN depending on a direction of movement of the shutter in relation to UCN gas. In case of emptying from the inner vessel shutters move into a vessel with UCN. There is mainly heating of UCN. In case of emptying from the annular vessel there is mainly UCN cooling since the shutter escapes from UCN flux. It is necessary to notice that this effect was observed experimentally. The peaks of heated neutrons are visible in the graphs of emptying process (Fig. 4) presented in [4,5]. Unfortunately, the effect has not been considered. It is neither discussed in work [1], nor in detailed work on this experiment [5]. These peaks are connected with UCN heating by shutters and are present only in case of emptying from the inner vessel. Unfortunately, it is not obviously possible to make a numerical estimation from the graphs. Therefore the given process was simulated.

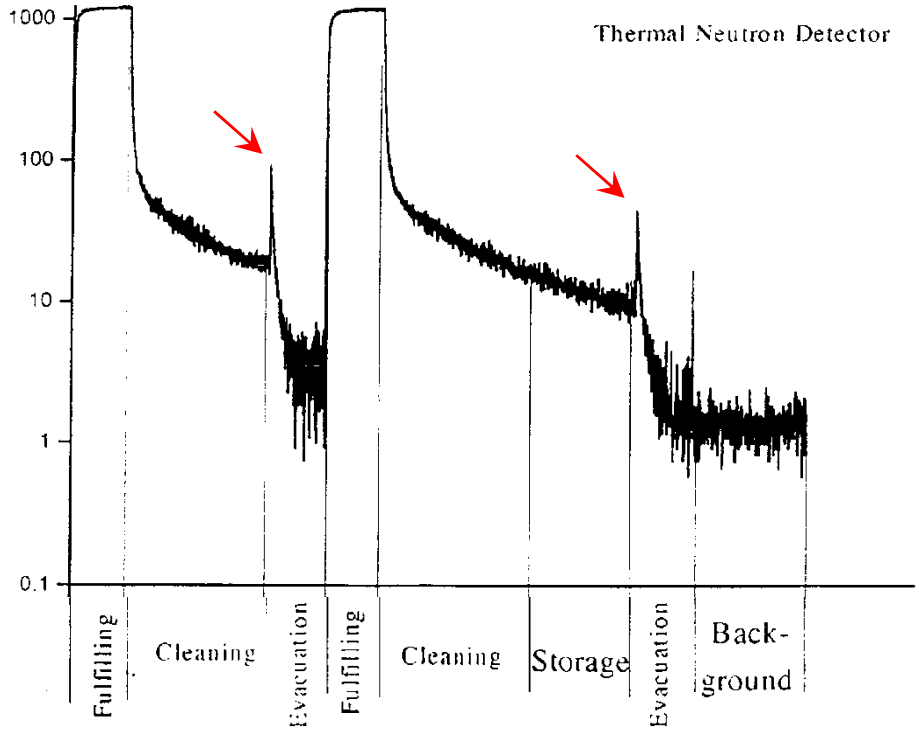

Fig. 4. Effect of heating of neutrons by the shutters.

When we work with the inner vessel the shutters 6 and 13 heat the trapped neutron spectrum after holding. Some part of UCN is lost due to this process. It gives lower value of $(N_i - N_f)$ and correspondingly higher value of $\lambda_{ie}$ for the inner vessel:

$$\lambda_{ie} = \frac{J\lambda}{(N_i - N_f)(1-\delta)} \frac{\varepsilon}{\varepsilon_{th}}, \qquad (15)$$

$$\lambda = \frac{1}{T} \ln \frac{N_i(1-\delta)}{N_f(1-\delta)}, \qquad (16)$$

where $\delta$ is part of neutrons heated by the shutters. The calculations were done with the shutter velocities ($v_{sh}$) of 1 and 2 m/s; the shutter course ($l_{sh}$) of 5, 10 and 15 cm. The arrow (2) in Fig. 3a shows the direction of a changed point position for the inner vessel after correction. It gives negative correction for measured value of neutron lifetime. The results of this simulation are shown in Fig. 3c. The correction for effect of UCN heating by shutters is –2.9 s for the shutter velocity of 1 m/s and the shutter course of 10 cm. As there are no detailed data on the shutters we cannot estimate uncertainty of this effect better than 2 s. Thus this correction is –2.9 s with uncertainty of initial data of 2 s.

**3. Effect of not equal thermal neutron detection efficiencies for different vessels.** Another obvious non-equivalence of measurements for different vessels is observed at thermal neutrons detection. The matter is that counters of thermal neutrons do not cover all external surface of the installation. They are absent at the installation end faces. For this reason processes of inelastic scattering occurring at the end faces of traps are registered with geometrical efficiency of about 50%. When neutrons are stored in the inner volume we have 2 end faces (on the left and on the right). But when neutrons are stored in the annular vessel there are 4 end faces (2 on the left and 2 on the right). In addition, the annular vessel is longer than the inner vessel and its end faces are more put forward. Unfortunately the value of this effect in work [5] is underestimated and wrongly considered with an opposite sign. For the estimation of non-equivalence effect in thermal neutrons the simulation of detection process has also been made.

The thermal detector efficiency is lower for the annular vessel because of 4 end faces. It gives lower value of $J$ and correspondingly lower value of $\lambda_{ie}$ for the annular vessel:

$$\lambda_{ie} = \frac{(J - \Delta J)\lambda}{(N_i - N_f)} \frac{\varepsilon}{\varepsilon_{th}}, \qquad (17)$$

where $\Delta J$ is number of not detected thermal neutrons for measurement with the annular vessel. We used mean values for the capture and scattering cross sections of materials of the setup from tables [6]. The simulation was done for the thermal neutron detector lengths ($L_{th}$) of 90, 100 and 110 cm. The arrow (3) in Fig. 3a shows the direction of changed position of point for the annular vessel after correction. It gives negative correction for the measured value of neutron lifetime. The results of this simulation are shown in Fig. 3d.

Geometrically the length of the detector is 100 cm, however its working area, apparently does not exceed 90 cm because of edge effects where devices of fastening of a thread are located. We choose the result of calculation for working length of the detector of 90 cm and for a case of capture and scattering of neutrons in an installation material. In this section we should notice that in work [5] effect of non-equivalence has been calculated, but the correction (+0.6 s) has appeared underestimated and with the wrong sign. Therefore we have to correct this error. Thus, the correction on effect of not equal thermal neutron detection efficiencies for different vessels is –2.1 s with uncertainty of initial data of 1 s.

Fig. 3a shows that each effect gives negative correction for the measured value of neutron lifetime. The summary table of corrections is shown in Table 1.

Table 1. MC correction on the neutron lifetime result of the experiment [1].

|  | correction, s | uncertainty, s |
|---|---|---|
| not full emptying of the inner vessel during cleaning while working with the annular vessel | 0 | 1 |
| effect of heating of neutrons by the shutters | –2.8 | 2 |
| effect of not equal thermal neutron detection efficiencies for different vessels | –2.1 | 1 |
| effect of not equal thermal neutron detection efficiencies for different vessels (correction in the experiment is +0.6 s) | –0.6 | |
| **total** | **–5.5** | **2.4** |

**Conclusion**

We assume that after taking into account MC correction and uncertainty the result of work [1] for neutron lifetime could be $879.9 \pm 0.9_{stat} \pm 2.4_{syst}$ s. The resulting corrected value for the neutron lifetime is in agreement with the result $878.5 \pm 0.8$ s of the work [2].

The authors are grateful to V.I. Morozov and L.N. Bondarenko for giving the information on the geometry of the experimental setup and critical remarks. The calculations were done at computing clusters: PNPI ITAD cluster, PNPI PC Farm. The given investigation has been supported by the Russian Foundation for Basic Research, projects no. 08-02-01052-a, 10-02-00217-a, 10-02-00224-a. The work has been supported by Federal Agency of Education, the state contracts no. P2427, P2500, P2540. The work has been supported by Federal Agency of Science and Innovations, the state contract no. 02.740.11.0532.